\documentstyle[epsf,axodraw]{elsart}
\journal{Physics Letters B}

\newcommand{\bsg}{b\to s\gamma }
\newcommand{\tb}{ \tan \beta}

\newcommand{\mrm}[1]{\;\mbox{\rm #1}}
\newcommand{\beq}{\begin{equation}}
\newcommand{\eeq}{\end{equation}}
\newcommand{\nn}{\nonumber}
\newcommand{\bea}{\begin{eqnarray}}
\newcommand{\eea}{\end{eqnarray}}

\newcommand{\rfn}[1]{(\ref{#1})}
\newcommand{\Eq}[1]{Eq.(\ref{#1})}

\newcommand{\ea}{{\it et al.}}

\newcommand{\eg}{{\it e.g. }}

\newcommand{\np}[1]{Nucl. Phys. {\bf #1}}
\newcommand{\plt}[1]{Phys. Lett. {\bf #1}}
\newcommand{\pr}[1]{Phys. Rev. {\bf #1}}
\newcommand{\prlt}[1]{Phys. Rev. Lett. {\bf #1}}
\newcommand{\zp}[1]{Z. Phys. {\bf #1}}

\newcommand{\prep}[1]{Phys. Rep. {\bf #1}}

\def\gsim{\mathrel{\vcenter{\hbox{$>$}\nointerlineskip\hbox{$\sim$}}}}

\begin{document}

\begin{frontmatter}

\begin{flushright} UCRHEP-T280\
\end{flushright}

\title{\bf Relaxing $b\to s \gamma$ Constraints on the Supersymmetric 
Particle Mass Spectrum}
\author[a1]{ Ernest Ma} and  
\author[a1,a2]{Martti Raidal}
\address[a1]{Department of Physics, University of California, Riverside,
CA 92521, U.S.A.}
\address[a2]{National Institute of Chemical Physics and Biophysics, 
R\"avala 10, 10143 Tallinn, Estonia }

\begin{abstract}
We consider the radiative decay $b \to s \gamma$ in a supersymmetric 
extension of the standard model of particle interactions, where the 
$b$-quark mass is entirely radiative in origin.  This is accomplished by 
the presence of nonholomorphic soft supersymmetry breaking terms in the 
Lagrangian.  As a result, the contributions to the $b \to s \gamma$ amplitude 
from the charged Higgs boson and the charginos/neutralinos are suppressed by 
$1/\tan^2\beta$ and ${\cal O}(\alpha/\alpha_s)$ respectively, allowing these 
particles to be lighter than in the usual supersymmetric model.  Their 
radiatively generated couplings differ from the usual tree-level ones and 
change the collider phenomenology drastically.  We also study how this 
scenario may be embedded into a larger framework, such as supersymmetric 
SU(5) grand unification.
\end{abstract}

\end{frontmatter}


{\bf 1.} {\it \underline{Introduction}.} 
The minimal supersymmetric standard model (MSSM) \cite{NHK} is one of the
most popular extensions of the standard model (SM). In recent years, both 
theorists and experimentalists have devoted enormous amounts of time to study 
its predictions.  While superpartner masses are expected to be below 1 TeV 
in some scenarios, there is actually a lot of uncertainty regarding 
the soft supersymmetry (SUSY) breaking sector of the theory.
In the most general case, the MSSM contains more than one hundred free 
parameters.  Nevertheless, present collider data as well as low-energy 
experiments are starting to place nontrivial constraints on the 
supersymmetric particle mass spectrum.

One of the processes known to put stringent constraints on new physics
is the radiative decay $\bsg$ with a branching ratio experimentally 
determined to be in the range \cite{cleo}

\begin{equation}
    2 \times 10^{-4} 
  <  {\rm BR}(B\to X_s\gamma)
  < 4.5\times 10^{-4}. 
\label{cleoband}
\end{equation}
This result agrees with the SM prediction.  On the other hand, in the MSSM 
framework, the decay $\bsg$ receives large additional contributions from 
charged Higgs-boson, chargino, neutralino, and gluino loops \cite{early}. 
The leading-order quantum-chromodynamics (LO QCD) corrections to $\bsg$ 
are known in the MSSM for arbitrary flavour structures \cite{early,bghw}, 
while the next-to-leading-order (NLO) analyses have been performed only for 
specific scenarios \cite{qcd}.  The charged Higgs-boson contribution 
always adds constructively to that of the SM (i.e. the W-boson contribution). 
The magnitudes of chargino and neutralino contributions depend strongly on 
$\tb$.  For large values of $\tb$, the chargino contribution becomes 
the dominant one.  In that case, its sign is determined by that of the $\mu$ 
parameter and can add constructively or destructively. 
Despite its strong interaction, the gluino contribution is generally the 
subdominant one.  Its sign depends on the details of the flavour structure 
of the model \cite{bghw}, \eg in SUSY SU(5) with a radiatively driven mass 
spectrum, it adds constructively to the chargino contribution \cite{bhr}.

An important issue for understanding the $\bsg$ constraints on SUSY models, 
recently reemphasized in Ref. \cite{cmw}, is the proper inclusion of loop 
corrections to the $b$-quark Yukawa coupling \cite{hall} when calculating 
the $\bsg$ rate.  These are enhanced for large $\tb$ and their sign is also 
determined by the $\mu$ parameter.  This implies a strong correlation between 
the values of the $b$-quark Yukawa coupling, SUSY model parameters, and the 
$\bsg$ constraints. 

The gauge-coupling unification in the MSSM strongly suggests that there is 
a grand unified theory (GUT) above the unification scale $M_{GUT}\sim 2 \times 
10^{16}$ GeV, such as SU(5) or SO(10) which allows tau-bottom or 
tau-bottom-top Yukawa coupling unification respectively.  However, successful 
Yukawa unification \cite{yuk} is achieved only for one sign of the $\mu$ 
parameter, $sign(\mu)=-$ in our convention, and for very large values of 
$\tb$, say $\sim 30-50$ for SU(5) and $\sim 50$ for SO(10) \cite{bdqt}. 
For $sign(\mu)=-$, all dominant contributions to $\bsg$ add constructively, 
implying thus very strong constraints on these SUSY scenarios.  Typically, 
the SUSY mass scale must exceed 1 TeV to be consistent with the bound given 
by \Eq{cleoband} \cite{bdqt}.  The charged Higgs-boson mass is forced to be 
large in this case, typically above the reach of the Tevatron at Fermilab 
as well as that of the Large Hadron Collider (LHC) at CERN.  In more general 
scenarios, $sign(\mu)=+$ allows cancellations between different terms.  
However for large $\tb$, the cancellation can happen only for a restricted 
part of the parameter space and stringent constraints may still be in force. 
In conclusion, the $\bsg$ bound of \Eq{cleoband} implies strong constraints 
on MSSM parameters in general, and on GUT scenarios in particular.

There are some proposals to satisfy the $\bsg$ constraints which still allow 
light sparticle masses accessible at future colliders.  In Ref. \cite{b2}, it 
has been pointed out in the context of SO(10) GUT that with $m,\,A\gg 
M_{1/2},$ where $m$ and $A$ denote generically common masses of matter 
multiplets and soft $A$ terms respectively, and $M_{1/2}$ is the common 
gaugino mass, there is a part of the parameter space for which the $\bsg$ 
rate is in the range given by \Eq{cleoband}.  In that case, the scalar 
superpartner masses are very large and suppress the new SUSY contributions 
to $\bsg$ whereas the wino masses can still be light.  This scenario implies 
that the only discoverable SUSY particles are light charginos and neutralinos.
The new Higgs bosons are heavy and the collider phenomenology discussed in 
Ref. \cite{cgnw} is not allowed.

The authors of Ref. \cite{bghw,cmw} argue instead that there might be
new flavour violation present in the squark sector of the model which
modifies and enhances the gluino contribution which then cancels the 
other large SUSY contributions. This scenario requires the introduction of 
new unknown flavour physics in general SUSY models and is not realized 
in GUTs \cite{bhr}. Also, the cancellation of two large terms cannot be 
considered a natural solution.

The purpose of this letter is to show that the dominant SUSY contributions
to $\bsg$ may be naturally suppressed if the SUSY radiative corrections 
to the $b$-quark Yukawa coupling are large.  We consider the limit of 
vanishingly small down-quark Yukawa couplings so that the corresponding 
quark masses are generated radiatively \cite{radmass,bfpt}.  Making the usual 
assumption that the trilinear $A$ terms are proportional to Yukawa couplings, 
we are forced to introduce the nonholomorphic $A'$ terms \cite{bfpt,hr,jj1,m} 
to give a correct mass to the $b$-quark.  After all, the soft $A'$ terms 
should be included in the complete SUSY Lagrangian on general grounds.  We 
find that in this scenario, the charged Higgs-boson and the dominant chargino 
contributions to $\bsg$ rate are suppressed by $1/\tan^2\beta$ and ${\cal O}
(\alpha/\alpha_s)$ respectively. Therefore, the soft, radiatively induced 
couplings of these particles change the Tevatron and LHC phenomenology 
considerably by reducing their production rates. This 
scenario can also be embedded into a GUT framework, such as SUSY SU(5), 
removing the stringent constraint on the MSSM parameter space coming from 
Yukawa unification.  We discuss the sparticle mass spectrum in that case.

{\bf 2.} {\it \underline{Proposed model and the radiative decay $\bsg$}.} 
In the following we work with the usual particle content of the MSSM 
\cite{NHK}.  However, we assume 
that the Yukawa coupling matrix in the $(d,s,b)$ quark sector is vanishing, 
i.e. $f_{d_{ij}}=0.$  In this case, the relevant MSSM superpotential $W$ 
for quark and Higgs left-chiral superfields is  
\beq
\label{mssmw}
W=  Q_i (f_{u_{ij}}) U^{c}_j H_2  - \mu H_1 H_2 .
\eeq
Further we make the usual assumption that the trilinear soft SUSY
breaking terms $a_{ij}$ have the same structure as the Yukawa coupling
matrices: $a_{ij}=A\cdot f_{ij}.$  Thus, neglecting leptons, the most general 
soft SUSY breaking terms in the MSSM are given by
\begin{eqnarray}
-{\cal L}_{\rm soft}&=& 
 \tilde{Q}_{i}^{\dagger}(m_{\tilde Q}^2)_{ij} \tilde{Q}_{j} 
+ \tilde{U}_{i}^{c\ast} (m_{\tilde U}^2)_{ij} \tilde{U}^{c}_{j}  
+ \tilde{D}_{i}^{c\ast} (m_{\tilde D}^2)_{ij} \tilde{D}^{c}_{j}  
 +m_{H_1}^2 H_1^\dagger H_1  
   +m_{H_2}^2 H_2^\dagger H_2  \nn\\
&& + \left(\tilde{Q}_{i} (A_{u} \cdot f_{u_{ij}})\tilde{U}^{c}_{j} H_2
  -\tilde{Q}_{i} (A'_{d} \cdot f_{u_{ij}})\tilde{D}^{c}_{j}(i\tau_2 H^\ast_2)
\right. \nn\\
&&  \left. 
 + B_{H} H_1H_2 
 + \frac12 M_1 \tilde{B} \tilde{B}
 + \frac12 M_2 \tilde{W}^a \tilde{W}^a 
 + \frac12 M_3 \tilde{g}^a \tilde{g}^a
   + h.c.\right) \,,
\label{soft}
\end{eqnarray}
where the $A'_d$ term is the nonholomorphic soft term and it does not cause 
quadratic divergences.  Therefore, it has been emphasized in Ref. \cite{hr}, 
and more recently in Ref. \cite{jj1}, that these terms should be included 
in the MSSM to study its low-energy phenomenology; their omission cannot 
be justified in the general context.  In the framework of high-energy 
physics, the nonholomorphic terms are generated, and not necessarily 
suppressed, in scenarios of spontaneous SUSY breaking such as those being 
mediated by supergravity \cite{m}.  In strongly coupled supersymmetric 
gauge theories, the nonholomorphic soft terms occur naturally \cite{ar}. 
Without understanding the real origin of SUSY breaking, we should keep the 
$A'$ terms in the general soft Lagrangian of the MSSM, such as \Eq{soft}.
Radiatively induced quark masses from the $A'$ terms have been
recently studied in Ref. \cite{bfpt}.

The $(u,c,t)$ quark masses come directly at tree level from the hard Yukawa
interaction in the superpotential \Eq{mssmw}.  However, the $(d,s,b)$ quark 
masses must be generated radiatively \cite{radmass,bfpt}. Taking only the 
dominant gluino-mediated contribution and neglecting intergenerational 
mixings, the one-loop soft bottom quark mass is given by 
\begin{equation}
 m_b =  - 2\frac{\alpha_s}{3 \pi} \,m_{\tilde{g}}\, m_t A'_b
\,I(m^{2}_{\tilde{b}_1}, m^{2}_{\tilde{b}_2}, m^2_{\tilde{g}})   \,.
\label{mb}
\end{equation}
Here the necessary chirality violation is due to the gluino mass
insertion and the term $m_t A'_b$ is the off-diagonal entry of 
the sbottom mass matrix.  The loop function $I(m_1^2,m_2^2,m_3^2)$ is 
given by \cite{hall}
\begin{equation}
 I(m_1^2,m_2^2,m_3^2)  =   - \frac{
 m_1^2 \,m_2^2 {\rm ln} ( m_1^2/ m_2^2 ) +
 m_2^2 \,m_3^2 {\rm ln} ( m_2^2/ m_3^2 )  +
 m_3^2 \,m_1^2 {\rm ln} ( m_3^2/ m_1^2 )
                                         }
{(m_1^2 -m_2^2)(m_2^2 -m_3^2)(m_3^2 -m_1^2)
          }\,.
\label{ifun}
\end{equation}
Obtaining the correct bottom quark mass via \Eq{mb} implies 
stringent constraints on the model parameters.

\begin{figure}
\begin{center}
\begin{picture}(270,110)(0,0)
\ArrowLine(0,10)(75,10)
\ArrowLine(195,10)(270,10)
\Line(270,10)(0,10)
\DashArrowArcn(135,10)(60,180,90){4}
\DashArrowArcn(135,10)(60,90,0){4}
\DashLine(135,70)(135,110){4}
\Text(135,0)[]{$\tilde g$}
\Text(37,0)[]{$ b_R$}
\Text(235,0)[]{$ t_L$}
\Text(155,97)[]{$H^+$}
\Text(75,60)[]{$\tilde b_R$}
\Text(195,60)[]{$\tilde t_L$}
\Text(135,10)[]{$\otimes$}
\end{picture}
\end{center}
\caption{Feynman diagram giving rise to the radiative 
$\bar t b H^+$ coupling. \vspace*{0.5cm}}
\label{fig:htb}
\end{figure}
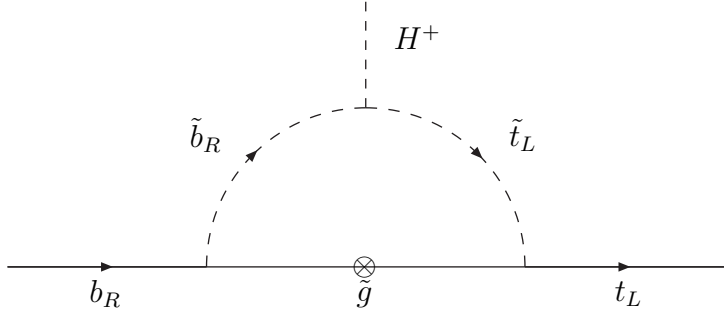

As the tree-level bottom Yukawa term is missing in the superpotential
\Eq{mssmw}, the $\bar t_L b_R H^+$ coupling is induced radiatively
by the diagram depicted in Fig. \ref{fig:htb}. The induced soft Lagrangian
at the one-loop level can be expressed as
\beq
{\cal L}^{\rm rad}_{\rm Yukawa}=- \frac{m_b}{v\tb} \bar t
\left[
r_1 P_L + r_2 P_R
\right]b H^+ + h.c. \,,
\label{radyuk}
\eeq
where the form factors $r_1$ and $r_2$ are given by
\bea
r_1 &=&
\frac{\sin2\theta_{\tilde t}\sin2\theta_{\tilde b}}
{4I(m^{2}_{\tilde{b}_1}, m^{2}_{\tilde{b}_2}, m^2_{\tilde{g}})}
\left[ 
C_0( 0,0,m_H^2; m_{\tilde{t}_1}^2,m^2_{\tilde{g}},m_{\tilde{b}_1}^2 ) 
 +
C_0( 0,0,m_H^2; m_{\tilde{t}_2}^2,m^2_{\tilde{g}},m_{\tilde{b}_2}^2 ) 
\right.\nn\\
&&  \qquad\qquad\qquad\quad - \left.
C_0( 0,0,m_H^2; m_{\tilde{t}_1}^2,m^2_{\tilde{g}},m_{\tilde{b}_2}^2 ) 
 -
C_0( 0,0,m_H^2; m_{\tilde{t}_2}^2,m^2_{\tilde{g}},m_{\tilde{b}_1}^2 ) 
\right]\,, \\
r_2 &=&
\frac{1}{I(m^{2}_{\tilde{b}_1}, m^{2}_{\tilde{b}_2}, m^2_{\tilde{g}})}
\left[ 
\cos^2\theta_{\tilde t}\sin^2\theta_{\tilde b}
C_0( 0,0,m_H^2; m_{\tilde{t}_1}^2,m^2_{\tilde{g}},m_{\tilde{b}_1}^2 ) 
\right. \nn\\
&& \qquad\qquad\qquad\quad +
\sin^2\theta_{\tilde t}\cos^2\theta_{\tilde b}
C_0( 0,0,m_H^2; m_{\tilde{t}_2}^2,m^2_{\tilde{g}},m_{\tilde{b}_2}^2 ) 
\nn\\
&&  \qquad\qquad\qquad\quad +
\cos^2\theta_{\tilde t}\cos^2\theta_{\tilde b}
C_0( 0,0,m_H^2; m_{\tilde{t}_1}^2,m^2_{\tilde{g}},m_{\tilde{b}_2}^2 ) 
\nn\\
&&  \qquad\qquad\qquad\quad + \left.
\sin^2\theta_{\tilde t}\sin^2\theta_{\tilde b}
C_0( 0,0,m_H^2; m_{\tilde{t}_2}^2,m^2_{\tilde{g}},m_{\tilde{b}_1}^2 ) 
\right] \,.
\eea
Here the three-point functions $C_0(0,0,m_H^2;m_1^2, m_2^2, m_3^2)$ 
are defined in \cite{bfpt} and we do not present them here.
In the limit $m_H\to 0,$ which is a good approximation if 
$m_H\ll m_{\tilde{g}},\,m_{\tilde{b}},\,m_{\tilde{t}}$ the 
three-point functions become
$C_0(0,0,0;m_1^2, m_2^2, m_3^2)\to I(m_1^2, m_2^2, m_3^2).$

\begin{figure}[t]
\centerline{
\epsfxsize = 0.5\textwidth \epsffile{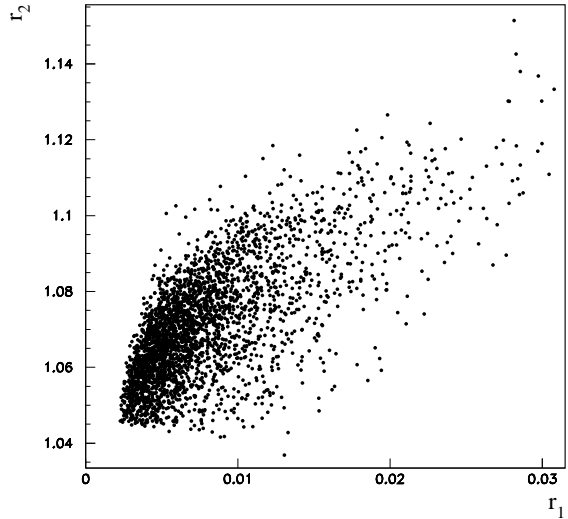} 
}
\caption{\it The scatter plot for the form factors $r_1$ and $r_2$
in the GUT scenario.
\vspace*{0.5cm}}
\label{fig:r}
\end{figure}

As with the usual hard tree-level coupling which we have omitted, the 
Lagrangian \Eq{radyuk} is also proportional to $m_b.$  However, the most 
important feature to notice is that the coupling is now suppressed by 
$\tb$ and not enhanced by it.   This is because the nonholomorphic $A'$ term 
couples to $H_2$ rather than to $H_1$ as in the usual case.  The induced 
couplings are momentum-dependent and are characterized by the form factors
$r_1$ and $r_2.$   If the couplings arise form the tree-level superpotential, 
then to lowest order, $r_1=0$ and $r_2=1.$  In our scenario, the deviation 
from these values is expected to be small, say a few percent.  Assuming that 
all stop and sbottom masses are degenerate as a first approximation, 
$m_{\tilde b_1}\approx m_{\tilde b_2}\approx m_{\tilde t_1}\approx 
m_{\tilde t_2},$ we have $r_1\to 0$ independently of the stop and sbottom 
mixing angles.  If the charged Higgs boson is lighter than the coloured 
sparticles, then we also have $r_2\to 1$ in this case. To show the realistic 
values of the form factors, we plot in Fig. \ref{fig:r} the values of $r_1$ 
against $r_2$ for the scenario (to be considered in detail in the next 
Section) where the SUSY mass spectrum is generated via renormalization-group 
running, assuming GUT conditions. Thus the interaction term proportional to 
$r_1$ in \Eq{radyuk} is just a small correction to the hard Yukawa term
arising from the superpotential \Eq{mssmw} and we neglect it in
further discussion. The second term in \Eq{radyuk} mimics the missing
hard term in the superpotential but is suppressed by $1/\tan^2\beta$
compared to it.

For quarks which obtain their masses radiatively, their higgsino couplings
are also generated radiatively.  The details for the neutral higgsinos are
given in Ref. \cite{bfpt}; similar arguments apply also for the 
charged higgsinos.  Without going into details, the most important
feature of the radiatively induced couplings of the 
right-handed $b$-quark to the higgsino $\tilde H$ and squark
is that it is induced by loops involving binos. 
Thus the radiatively induced couplings for $b_R$ are always 
suppressed by ${\cal O}(\alpha/\alpha_s)$ compared to the hard couplings.
We show below that this suppression factor also applies to the 
dominant chargino and neutralino contributions to the $\bsg$ amplitude.

The effective Hamiltonian for $b\to s\gamma$ in SUSY models 
can be expressed in two terms:
\beq
{\cal H}_{eff}= {\cal H}_{eff}^{CKM} + {\cal H}_{eff}^{\tilde g}\,,
\eeq
where 
\begin{equation}
 {\cal H}_{eff}^{CKM} = 
 - \frac{4 G_F}{\sqrt{2}} V_{tb}^{\phantom{\ast}} V_{ts}^\ast
  \sum_i C_i(\mu) {\cal O}_i(\mu) \, 
\label{weffham}
\end{equation}
contains the SM as well as the charged Higgs-boson, chargino, and neutralino
contributions with the same flavour structure as in the SM.
The explicit formulas including LO QCD corrections can be found  
in Ref. \cite{early}.
The gluino contribution ${\cal H}_{eff}^{\tilde g}$ may exhibit
in addition the new flavour violation present in general SUSY models
but absent in our scenario;
details including LO QCD corrections can be found in Ref. \cite{bghw}. 
The decay width of $b \to s \gamma$ can be written as
\begin{equation}
 \Gamma(b \to s \gamma) = 
 \frac{m_b^5 \, G_F^2 \, |V_{tb} V_{ts}^*|^2
 \, \alpha}{32 \pi^4} 
\left\vert C^{eff}_7  \right\vert^2  \,,
\label{c7}
\end{equation}
where $\left\vert C^{eff}_7  \right\vert^2 = 
\left\vert C_7 + C^{\tilde g}_7 \right\vert^2 +
\left\vert C^{'\tilde g}_7  \right\vert^2 \,. $
Here $C_7$ stands for the total contribution from the effective
Hamiltonian \Eq{weffham}, while $C^{\tilde g}_7$ and $C'^{\tilde g}_7$  
arise from gluino loops.
The present experimental result \Eq{cleoband} 
implies the allowed range:
\beq
 0.25 < |C^{eff}_7| < 0.375 \,.
\label{c7range}
\eeq

\begin{figure}
\begin{center}
\begin{picture}(270,110)(0,0)
\Line(260,10)(10,10)
\ArrowLine(65,10)(115,10)
\ArrowLine(15,10)(65,10)
\ArrowLine(205,10)(255,10)
\ArrowLine(150,10)(205,10)
\DashCArc(135,10)(70,0,180){5}
\GCirc(65,10){10}{0.3}
\Photon(0,90)(70,45){4}{5}
\Text(180,0)[]{$ t_R$}
\Text(95,0)[]{$ t_L$}
\Text(37,0)[]{$ b_R$}
\Text(235,0)[]{$ s_L$}
\Text(155,93)[]{$H^+$}
\Text(135,10)[]{$\otimes$}
\end{picture}
\end{center}
\caption{Charged Higgs-boson contribution to $\bsg$ due to the radiative
coupling. The blob is generated by the diagram in Fig. \ref{fig:htb} and
is proportional to $m_b.$ \vspace*{0.5cm}}
\label{fig:bsg}
\end{figure}
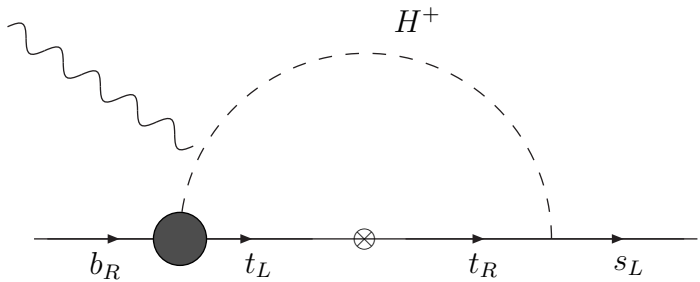

Let us now consider different SUSY contributions to $\bsg$ in our 
scenario. The charged Higgs-boson contribution is induced by two different
chiral structures: the SM-like one with the chirality flip in the 
external $b$-quark line which is induced by the hard top Yukawa
interactions, and the one with the chirality flip in the internal 
$t$-quark line which is induced by the radiative coupling \Eq{radyuk}. 
The latter diagram, presented in Fig. \ref{fig:bsg}, is
actually a two-loop diagram.  A rigorous calculation would require that it be 
performed in two loops.   However, since the coloured sparticles
are expected to be much heavier than the Higgs bosons, we can 
integrate them out at their mass scale and assume with a good
accuracy that the $H^+$ contribution is given by the one-loop
diagram in Fig. \ref{fig:bsg} with the radiative coupling \Eq{radyuk}.
In that case, the latter contribution to $C_7$, 
which is not suppressed by $\tb$
if the bottom Yukawa couplings are hard, becomes suppressed by 
$1/\tan^2\beta$ implying that the full  $H^+$ contribution to $C_7$
is suppressed by the same factor. Explicitly,
\beq
C_7^{H^+}(m_W)=\frac{1}{2} \frac{x_{tH}}{\tan^2\beta}
\left[ \frac{2}{3} f_1(x_{tH})  + f_2(x_{tH}) + r_2
\left(\frac{2}{3} f_3(x_{tH})  + f_4(x_{tH})\right) \right] \,,
\eeq
where $x_{tH}=M_{H^+}^2/m_t^2$ and the Inami-Lim type functions 
$f_i(x_{tH})$ can be found in Ref. \cite{early}.
This equation implies an important result: the mass bounds on the 
charged Higgs boson coming from the measurement of $\bsg$
can be relaxed and $H^+$ can be light.

The chargino and neutralino contributions to $C_7$ follow the same 
two chiral patterns discussed above. In this case the chirality
flip may occur in the internal  chargino/neutralino line implying
the large enhancement factors $m_{\tilde\chi_i}/m_b.$ These enhanced terms,
induced by the higgsinos,
dominate the  chargino/neutralino contributions. 
However, as we argued before, the relevant higgsino couplings are suppressed 
by ${\cal O}(\alpha/\alpha_s)$ compared to the case of hard bottom
Yukawa couplings. The exact calculation of $C_7^{\chi^+,\chi^0}$
involves two loops and is beyond the aim of this letter.
In our numerical examples below, we take the known MSSM expressions
for $C_7^{\chi^+,\chi^0}$ from Ref. \cite{early}
and suppress the dominant chirality flipping terms by  $\alpha/\alpha_s.$

Because the $b$-quark mass is generated radiatively by the gluino loop,
the gluino-mediated contribution to $\bsg$ is expected to be 
sizable.  However, in the absence of new large flavour violation
beyond the Cabibbo-Kobayashi-Maskawa (CKM) matrix, the gluino contribution 
is always subdominant compared with that of the SM or the charged Higgs-boson 
and chargino ones.  Nevertheless, in our scenario, the gluino contribution 
may become the largest SUSY contribution to $\bsg.$

To conclude this Section, in the scenario of radiatively induced
$(d,s,b)$ quark masses in the MSSM, the $\bsg$ rate does 
not impose serious constraints on the charged Higgs-boson, chargino, 
and neutralino masses.   This allows the possibility of their 
production at colliders but with drastically modified couplings, 
implying new phenomenology at experiments.

{\bf 3.} {\it \underline{Unification framework}.} 
Embedding the MSSM with the radiative $(d,s,b)$ quark masses into a 
unification framework is a well-motivated and appealing possibility.
This can naturally happen in SUSY SU(5) GUT because the up and down Yukawa
couplings are not unified in this model. This automatically 
solves all the constraints on the model parameters 
(see, \eg, Ref. \cite{bdqt}) 
coming from the prediction of $b-\tau$ Yukawa unification; 
they are simply vanishing. In addition, the constraints from $\bsg$ 
are practically removed, as shown in the previous Section.

Nevertheless the mass spectrum in such a version of the MSSM is stringently
constrained by the renormalization-group running of the model parameters, 
the requirement of radiative electroweak symmetry breaking, and most 
importantly, the requirement of generating a correct mass to the $b$ quark 
via \Eq{mb}.  We start the running of gauge and top Yukawa couplings
at $m_t$ using the two-loop MSSM renormalization group equations \cite{MV}.
The bottom and tau Yukawa couplings are taken to be vanishing. We 
identify the unification scale $M_{GUT}=2\cdot 10 ^{16}$ GeV by the meeting of 
$g_1$ and $g_2.$ At that scale we generate randomly the free parameters
of the model: the common gaugino mass $M_{1/2},$ 
common squark mass $m_{0},$ common Higgs mass $M_{H_0},$ 
common $A$ parameter $A_0,$ common $A'$ parameter $A'_0,$  
$\tb$, and $sign(\mu)$ in the following ranges:
\bea
100\, \mrm{GeV} < & M_{1/2} & < 1000 \,\mrm{GeV}\,, \nn\\
100\, \mrm{GeV} < & m_{0} & < 1000 \,\mrm{GeV}\,, \nn\\
100\, \mrm{GeV} < & M_{H_0} & < 1000 \,\mrm{GeV}\,, \nn\\
-2000\, \mrm{GeV} < & A_{0} & < 1000 \,\mrm{GeV}\,, \nn\\
-2000\, \mrm{GeV} < & A'_{0} & < 1000 \,\mrm{GeV}\,, \nn\\
3\,  < & \tb & < 60 \,.
\eea 
Note that $\tb$ is now a free parameter and is not constrained
by $b-\tau$ Yukawa unification. With these initial values we run the
model parameters to the weak scale, assume radiative symmetry-breaking
conditions and calculate the sparticle and Higgs-boson mass
matrices there. The renormalization-group equations for nonholomorphic
terms can be found in \cite{jj1}. 
We require that the branching ratio of $\bsg$ is in the
allowed range \rfn{cleoband} and that the radiative $b$-quark mass
\Eq{mb} is in the range $2.8 < m_b(m_Z) < 3.2.$ 
The scatter plots are almost independent of the sign of the $\mu$ parameter, 
our results are presented for $sign(\mu)=-.$

\begin{figure}[t]
\centerline{
\epsfxsize = 0.5\textwidth \epsffile{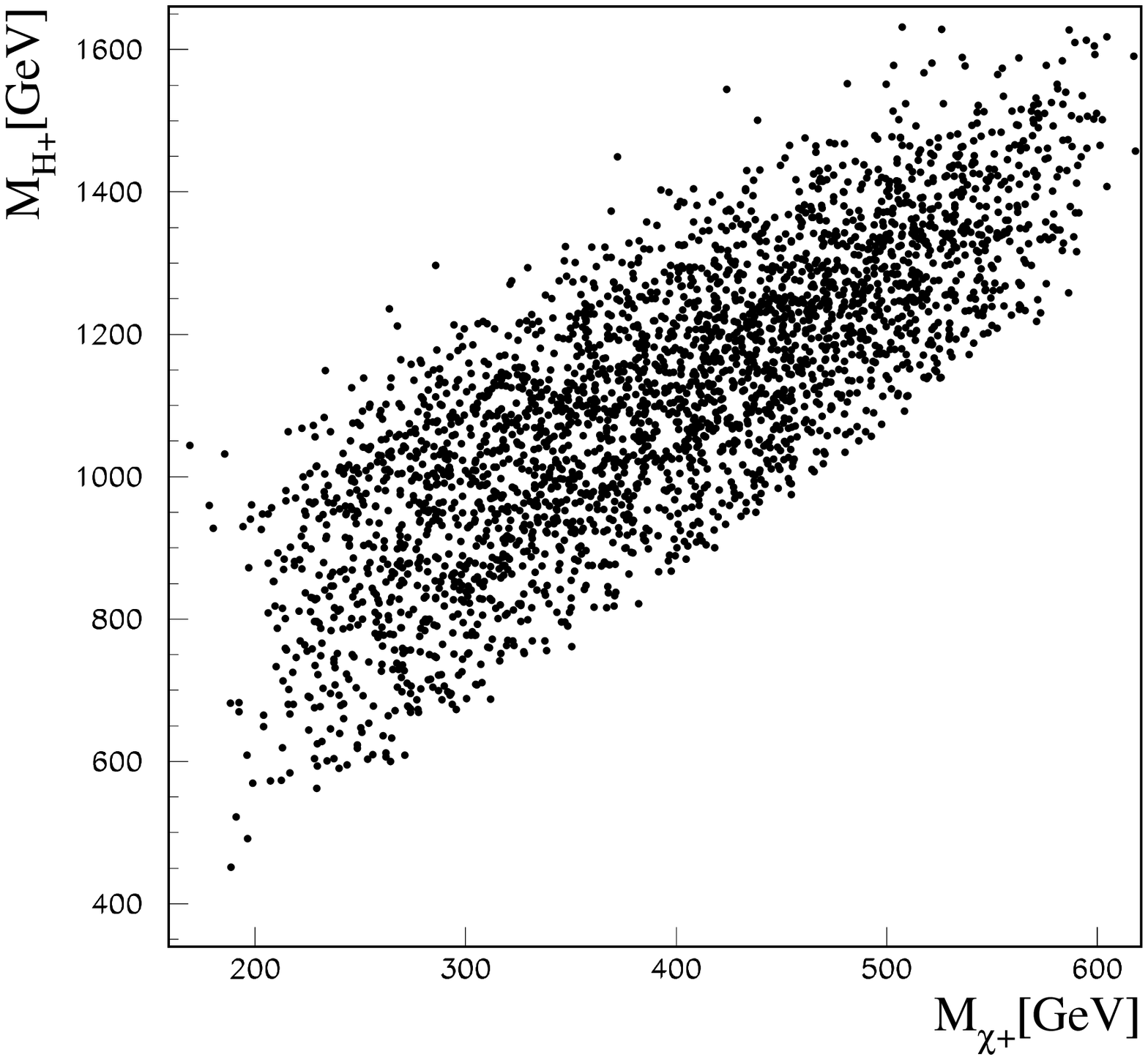} 
\hfill
\epsfxsize = 0.5\textwidth \epsffile{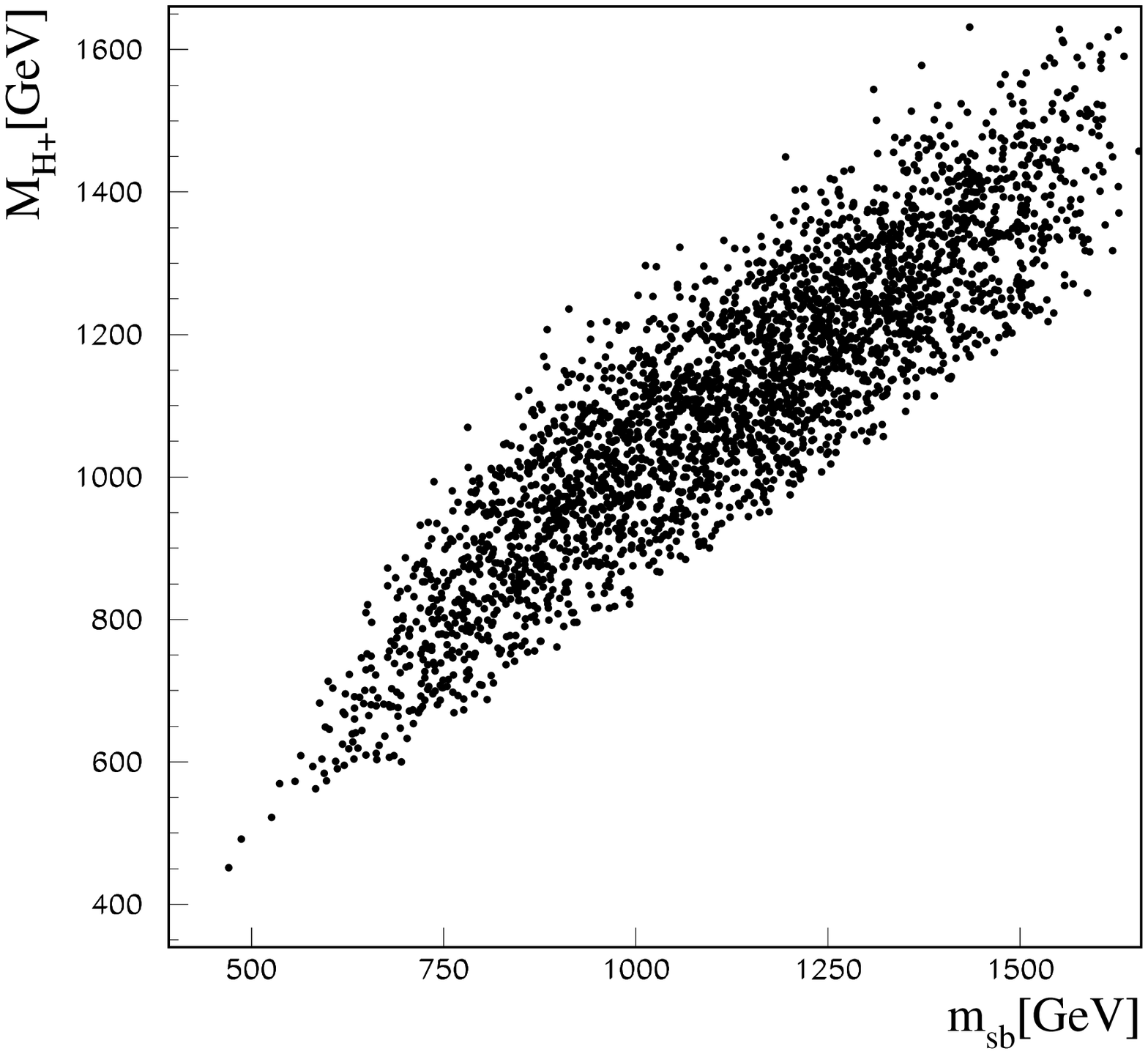}
}
\caption{\it  Scatter plots of the allowed values of the charged Higgs-boson 
mass $M_{H^+}$ against the lightest chargino mass $M_{\tilde\chi^+_1},$ 
and against the lightest bottom squark mass $m_{\tilde b_1}.$
\vspace*{0.5cm}}
\label{fig:mch}
\end{figure}

In Fig. \ref{fig:mch} we present the scatter plots of the allowed
values of the charged Higgs-boson mass $M_{H^+}$ against the lightest 
chargino mass $M_{\tilde\chi^+},$ 
and against the lightest bottom squark mass $m_{\tilde b_1}.$
Notice that in this scenario $H^+$ is bounded to be rather heavy,
$M_{H^+}\gsim 400$ GeV. This comes from the requirement of 
radiatively induced electroweak symmetry breaking. Because the 
$b$-quark Yukawa coupling is vanishing, the difference between the 
Higgs-boson mass parameters $M_{H_1}$ and $M_{H_2}$ is maximized.
 As their difference determines $m_A$, the charged
Higgs boson is naturally quite heavy.
However, the charginos can be light and be discovered at future colliders.

\begin{figure}[t]
\centerline{
\epsfxsize = 0.5\textwidth \epsffile{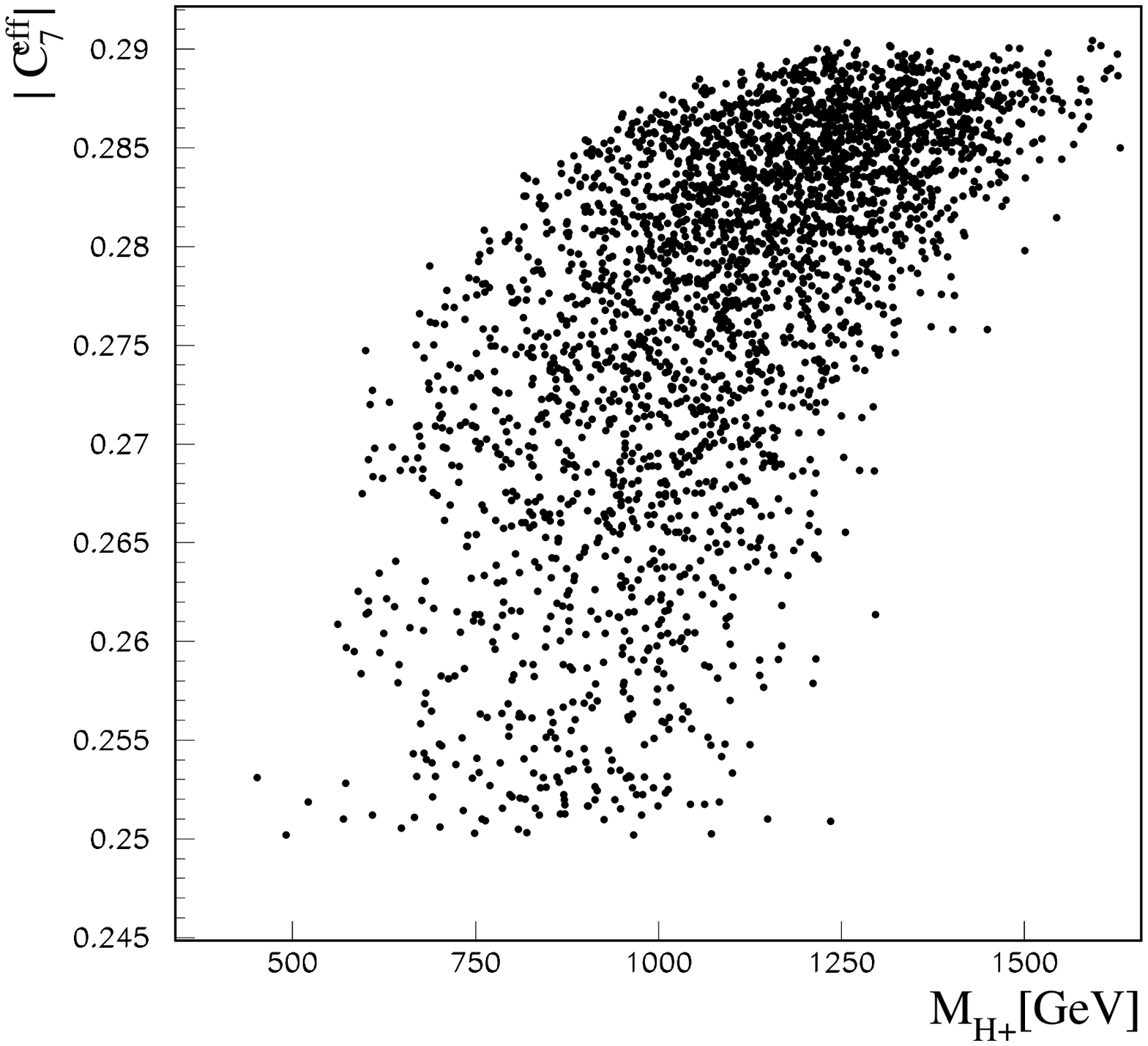} 
\hfill
\epsfxsize = 0.5\textwidth \epsffile{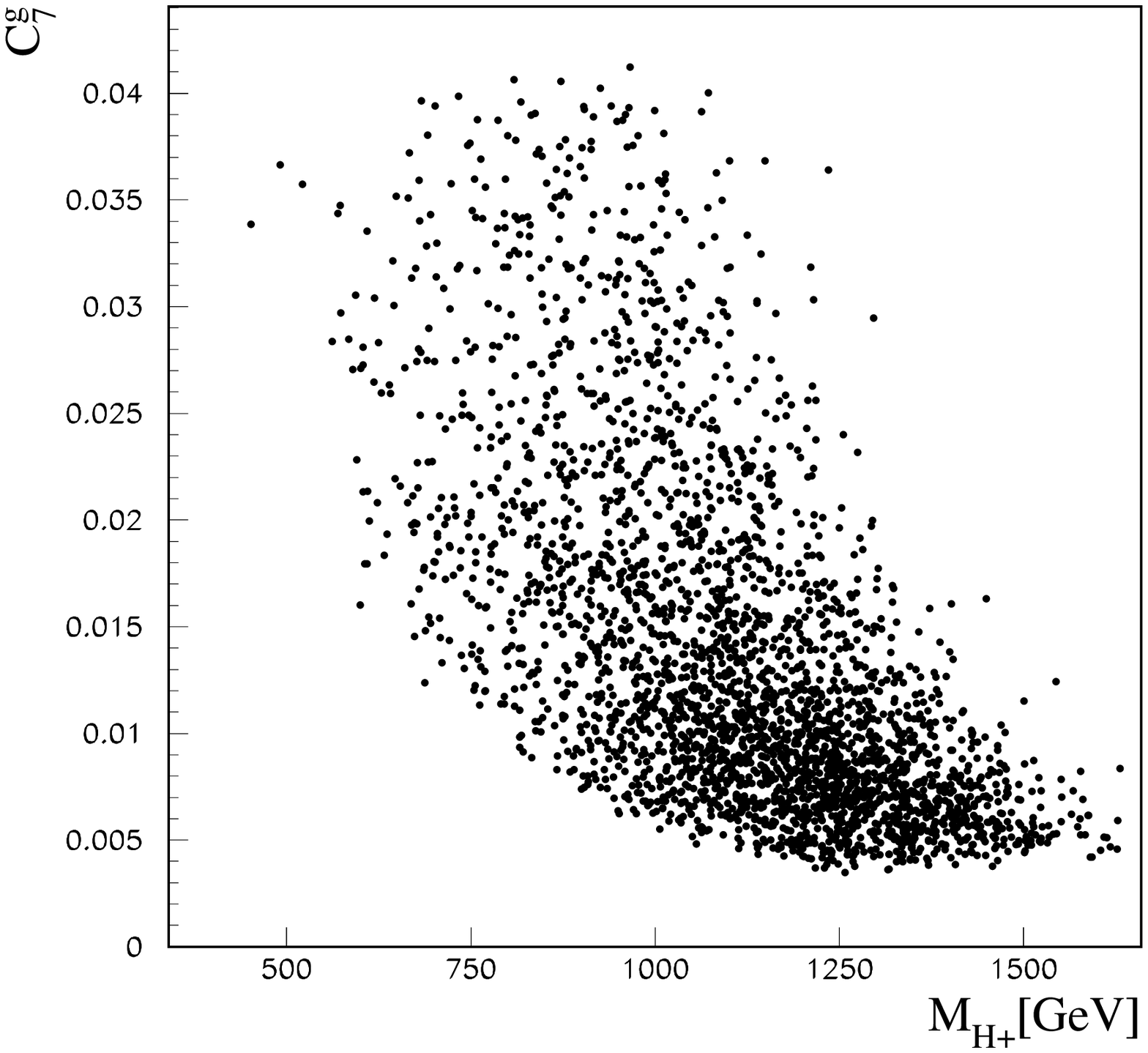}
}
\caption{\it Scatter plots of the total $|C_7^{eff}|$ and the 
dominant gluino contribution $C_7^{\tilde g}$
against the charged Higgs-boson mass $M_{H^+}.$
\vspace*{0.5cm}}
\label{fig:c7}
\end{figure}

It is interesting to see the SUSY contribution to $\bsg$ in this scenario.
In Fig. \ref{fig:c7} we plot the total value of $|C_7^{eff}|$ and the
gluino contribution $C_7^{\tilde g}$ (see \Eq{c7} for explanation)
against the charged Higgs-boson mass  $M_{H^+}.$ The deviation from the LO SM
value $C_7^{SM}=-0.29$ is small and, as follows from the 
figures,  dominated by the gluino contribution. Thus in this scenario, 
the MSSM mass spectrum is not constrained by $\bsg.$


{\bf 4.} {\it \underline{Conclusions}.} 
We have studied the decay $\bsg$ in the MSSM in the case the
$(d,s,b)$ quark masses are generated radiatively. The soft radiatively 
generated $b_R$ couplings to the charged Higgs boson and higgsino are 
suppressed by $1/\tan^2\beta$ and ${\cal O}(\alpha/\alpha_s)$ respectively. 
The dominant contributions
of these particles to $\bsg$ are suppressed by the same factors allowing
the existence of light $H^+$ and $\tilde\chi^+.$  Their production and 
decay processes at future colliders are changed drastically.

If this scenario is realized in the framework of GUTs, then 
the constraints from $b-\tau$ Yukawa unification as well as from
$\bsg$ are removed. Nevertheless, the lightest sparticles in that case
are binos and winos.


\begin{ack}
This work was supported in part by the U.~S.~Department of Energy
under Grant No. DE-FG03-94ER40837. 
\end{ack}

\end{document}